\documentclass[english]{article}
\usepackage[T1]{fontenc}
\usepackage[latin9]{inputenc}
\usepackage{amssymb}

\makeatletter
\newcommand{\lyxaddress}[1]{
\par {\raggedright #1
\vspace{1.4em}
\noindent\par}
}

\makeatother

\usepackage{babel}
\begin{document}

\title{\textbf{Quantum corrected non-thermal radiation spectrum from the
tunnelling mechanism}}

\author{\textbf{$^{1}$Subenoy Chakraborty, $^{2}$Subhajit Saha and $^{3,4,5}$Christian
Corda}}

\maketitle

\lyxaddress{\textbf{$^{1,2}$Department of Mathematics, Jadavpur University,
Kolkata 700032, West Bengal, India}}

\lyxaddress{\textbf{$^{3}$Dipartimento di Scienze, Sezione di Fisica,} \textbf{Scuola
Superiore di Studi Universitari e Ricerca \textquotedbl{}Santa Rita\textquotedbl{},
via San Nicola snc, 81049, San Pietro Infine (CE) Italy}}

\lyxaddress{\textbf{$^{4}$Austro-Ukrainian Institute for Science and Technology,
Institut for Theoretish Wiedner Hauptstrasse 8-10/136, A-1040, Wien,
Austria}}

\lyxaddress{\textbf{$^{5}$International Institute for Applicable Mathematics
\& Information Sciences (IIAMIS), Hyderabad (India) \& Udine (Italy)}}

\lyxaddress{\textbf{Email addresses: }\textbf{\emph{schakraborty.math@gmail.com,
subhajit1729@gmail.com, cordac.galilei@gmail.com}}}
\begin{abstract}
Tunnelling mechanism is today considered a popular and widely used
method in describing Hawking radiation. However, in relation to black
hole (BH) emission, this mechanism is mostly used to obtain the Hawking
temperature by comparing the probability of emission of an outgoing
particle with the Boltzmann factor. On the other hand, Banerjee and
Majhi reformulated the tunnelling framework deriving a black body
spectrum through the density matrix for the outgoing modes for both
the Bose-Einstein distribution and the Fermi-Dirac distribution. In
contrast, Parikh and Wilczek introduced a correction term performing
an exact calculation of the action for a tunnelling spherically symmetric
particle and, as a result, the probability of emission of an outgoing
particle corresponds to a non-strictly thermal radiation spectrum.
Recently, one of us (C. Corda) introduced a BH effective state and
was able to obtain a non-strictly black body spectrum from the tunnelling
mechanism corresponding to the probability of emission of an outgoing
particle found by Parikh and Wilczek. The present work introduces
the quantum corrected effective temperature and the corresponding
quantum corrected effective metric is written using Hawking's periodicity
arguments. Thus, we obtain further corrections to the non-strictly
thermal BH radiation spectrum as the final distributions take into
account both the BH dynamical geometry during the emission of the
particle and the quantum corrections to the semiclassical Hawking
temperature.
\end{abstract}


Considering Hawking radiation \cite{key-1} in the tunnelling approach
\cite{key-2}-\cite{key-7}, \cite{key-25}-\cite{key-27} the particle
creation mechanism caused by the vacuum fluctuations near the BH horizon
works as follows. A virtual particle pair is created just inside the
horizon and the virtual particle with positive energy can tunnel out
the BH horizon as a real particle. Otherwise, the virtual particle
pair is created just outside the horizon and the negative energy particle
can tunnel inwards. Thus, for both the possibilities, the particle
with negative energy is absorbed by the BH and as a result the mass
of the BH decreases. The flow of positive energy particles towards
infinity is considered as Hawking radiation. Earlier, this approach
was limited to obtain only the Hawking temperature through a comparison
of the probability of emission of an outgoing particle with the Boltzmann
factor rather than the actual radiation spectrum with the correspondent
distributions. This problem was formally addressed by Banerjee and
Majhi \cite{key-7}. By a novel formulation of the tunnelling formalism,
they were able to directly reproduce the black body spectrum for either
bosons or fermions from a BH with standard Hawking temperature. However,
considering contributions beyond semiclassical approximation in the
tunnelling process, Parikh and Wilczek \cite{key-2,key-3} found a
probability of emission compatible with a non-thermal spectrum of
the radiation from BH. This non precisely thermal character of the
spectrum is important to resolve the information loss paradox of BH
evaporation \cite{key-8} because arguments that information is lost
during hole's evaporation partially rely on the assumption of strict
thermal behavior of the radiation spectrum \cite{key-3,key-8}. Interesting
approaches to resolve the BH information puzzle have been recently
proposed in \cite{key-9,key-10}

The basic difference between the works \cite{key-2,key-3} and the
work \cite{key-7} is consideration or non-consideration of the energy
conservation. As a result, there will be a dynamical \cite{key-2,key-3}
or static \cite{key-7} BH geometry. In fact, due to conservation
of energy, in \cite{key-2,key-3} the BH horizon contracts during
the radiation process which deviates from the perfect black body spectrum.
This non-thermal spectrum has profound implications for realizing
the underlying quantum gravity theory. In the language of the tunnelling
mechanism, a trajectory in imaginary or complex time joins two separated
classical turning points \cite{key-3}. The key point is that the
forbidden region traversed by the emitting particle has a \emph{finite}
size \cite{key-3} from $r=r_{initial}$ to $r=r_{final}$ ($r_{initial}$
is the radius of the horizon of the BH initially and $r_{final}$
is the radius of the horizon of the BH after particle emission). This
finite size implies a discrete nature of the tunnelling mechanism,
which is characterized by the physical state before the emission of
the particle and that after the emission of the particle. As a result,
the radiation spectrum is also discrete \cite{key-10,key-11}. Consequently,
particle emission can be interpreted like a quantum transition of
frequency $\omega$ between the two discrete states \cite{key-10,key-11}.
It is the particle itself which generates a tunnel through the horizon
\cite{key-3,key-10,key-11} having finite size. In thermal spectrum,
the tunnelling points have zero separation, so there is no clear trajectory
because there is no barrier \cite{key-3,key-10,key-11}. 

In Planck units ($G=c=k_{B}=\hbar=\frac{1}{4\pi\epsilon_{0}}=1$ ),
the strictly thermal tunnelling probability is given by \cite{key-1,key-2,key-3}
\begin{equation}
\Gamma\sim\exp\left(-\frac{\omega}{T_{H}}\right),
\end{equation}
where $T_{H}=\frac{1}{8\pi M}$ is the Hawking temperature and $\omega$
is the energy-frequency of the emitted radiation. However, considering
contributions beyond semiclassical approximation and taking into account
the conservation of energy, Parikh and Wilczek reformulate the tunnelling
probability as \cite{key-2,key-3} 
\begin{equation}
\Gamma\sim\exp\left[-\frac{\omega}{T_{H}}\left(1-\frac{\omega}{2M}\right)\right].
\end{equation}
This non-thermal spectrum enables the introduction of an intriguing
way to consider the BH dynamical geometry through the \emph{BH effective
state. }In fact, one introduces the \emph{effective temperature }as
\cite{key-10}-\cite{key-13} 
\begin{equation}
T_{E}(\omega)\equiv\frac{2M}{2M-\omega}T_{H}=\frac{1}{4\pi(2M-\omega)},
\end{equation}
which permits to rewrite the probability of emission (2) in Boltzmann-Hawking
form as \cite{key-10}-\cite{key-13} 
\begin{equation}
\Gamma\sim\exp[-\beta_{E}(\omega)\omega]=\exp(-\frac{\omega}{T_{E}(\omega)}),\label{eq: Corda Probability}
\end{equation}
where the effective Boltzmann factor takes the form \cite{key-10}-\cite{key-13}
\begin{equation}
\beta_{E}(\omega)\equiv\frac{1}{T_{E}(\omega)}.
\end{equation}
One interpretes the effective temperature as the temperature of a
black body emiting the same total amount of radiation \cite{key-10}-\cite{key-13}.
{\bf Hence Hawking temperature is replaced by the effective temperature in
the expression for the probability of emission. It should be noted that
this notion of effective temperature has already been introduced in the
literature for the Schwarzschild BH \cite{key-12, key-13}, for the Kerr
BH \cite{key-19} and for the Reissner-Nordstrom BH \cite{key-20}. Further,
the ratio $\frac{T_E(\omega)}{T_H}=\frac{2M}{2M-\omega}$ characterize the
deviation of the radiation spectrum of a BH from the strictly thermal 
feature \cite{key-10, key-11, key-12, key-13}. Also, the tunnelling 
approach of Parikh and Wilczek shows the probability of emission
of Hawking quanta [see Eq. (2)] is non-thermal in nature (i.e., BH does
not emit like a perfect black body). Moreover, due to perfect black body
character of Bose-Einstein and Fermi-Dirac distributions, it is natural
to have deviations from these distributions in case of the above effective
temperature. Thus in analogy to BH, the effective temperature of a body
(say star) can be defined as the temperature of a black body that would 
emit the same total amount of electromagnetic radiation \cite{key-10, key-14}.
So one can consider this effective temperature and the bolometric luminosity
as the two fundamental physical parameters to identify a star on the Hertzsprung-Russel diagram. It is worthy to mention here that both the above two physical parameters however depend on the chemical composition of the star \cite{key-10,
key-11, key-12, key-13, key-14}.}

Further, in analogy with the effective temperature, one can define
the \emph{effective mass} and the \emph{effective horizon} radius
as \cite{key-10}-\cite{key-13}
\begin{equation}
M_{E}=M-\frac{\omega}{2}\qquad and\qquad r_{E}=2M_{E}=2M-\omega.
\end{equation}
Note that these effective quantities are nothing but the average value
of the corresponding quantities before (initial) and after (final)
the particle emission (i.e., $M_{i}=M$, $M_{f}=M-\omega$; $r_{i}=2M_{i}$
and $r_{f}=2M_{f}$). Accordingly, $T_{E}$ is the inverse of the
average value of the inverses of the initial and final Hawking temperatures
\cite{key-10}-\cite{key-13}. Hence, there is a discrete character
(in time) of the Hawking temperature. Thus, the effective temperature
may be interpreted as the Hawking temperature \emph{during} the emission
of the particle \cite{key-10}-\cite{key-13}. 

\noindent Following \cite{key-11} one can use Hawking's periodicity
argument \cite{key-11,key-23,key-24} to obtain the \emph{effective
Schwarzschild line element} 
\begin{equation}
ds_{E}^{2}=-(1-\frac{2M_{E}}{r})dt^{2}+\frac{dr^{2}}{1-\frac{2M_{E}}{r}}+r^{2}(\sin^{2}\theta d\varphi^{2}+d\theta^{2}),\label{eq: Hilbert effective}
\end{equation}
which takes into account the BH \emph{dynamical} geometry during the
emission of the particle. 

Recently, one of us (C. Corda) introduced the above discussed BH effective
state \cite{key-10}-\cite{key-13} and was able to obtain a non-strictly
black body spectrum from the tunnelling mechanism corresponding to
the probability of emission of an outgoing particle found by Parikh
and Wilczek \cite{key-11}. The final non-strictly thermal distributions
which take into account the BH dynamical geometry are \cite{key-10,key-11}
\begin{equation}
\begin{array}{c}
<n>_{boson}=\frac{1}{\exp\left[4\pi\left(2M-\omega\right)\omega\right]-1}\\
\\
<n>_{fermion}=\frac{1}{\exp\left[4\pi\left(2M-\omega\right)\omega\right]+1}.
\end{array}\label{eq: final distributions-1}
\end{equation}
Now, we further modify the effective temperature by incorporating
the quantum corrections to the semiclassical Hawking temperature discussed
in \cite{key-4}. As a result, the quantum physics of BHs will be
further modified. Banerjee and Majhi \cite{key-4} have formulated
the quantum corrected Hawking temperature using the Hamilton-Jacobi
method \cite{key-15} beyond semiclassical approximation. According
to them \cite{key-4}, the quantum corrected Hawking temperature (termed
as \emph{modified Hawking temperature}) is given by 
\begin{equation}
T_{H}^{(m)}=\left[1+\Sigma_{i}\frac{\beta_{i}}{M^{2i}}\right]^{-1}T_{H},
\end{equation}
where the $\beta_{i}$ are dimensionless constant parameters. However,
if these parameters are chosen as powers of a single parameter $\alpha$,
then in compact form \cite{key-4} 
\begin{equation}
T_{H}^{(m)}=\left(1-\frac{\alpha}{M^{2}}\right)T_{H}.
\end{equation}
This modified Hawking temperature is very similar in form to the temperature
correction in the context of one-loop back reaction effects \cite{key-16,key-17}
in the spacetime with $\alpha$ related to the trace anomaly \cite{key-18}.
Further, using conformal field theory, if one considers one-loop quantum
correction to the surface gravity for Schwarzschild BH then $\alpha$
has the expression \cite{key-4} 
\begin{equation}
\alpha=-\frac{1}{360\pi}\left(-N_{0}-\frac{7}{4}N_{\frac{1}{2}}+13N_{1}+\frac{233}{4}N_{\frac{3}{2}}-212N_{2}\right),
\end{equation}
where $N_{s}$ denotes the number of field with spin $s$. Also considering
two-loop back reaction effects in the spacetime, the quantum corrected
Hawking temperature becomes \cite{key-4} 
\begin{equation}
T_{H}^{(m)}=\left[1-\frac{\alpha}{M^{2}}-\frac{\gamma}{M^{4}}\right]T_{H},\label{eq: 12}
\end{equation}
where second loop contributions are related to the dimensionless parameter
$\gamma$. Thus, it is possible to incorporate higher loop quantum
corrections by proper choices of the $\beta_{i}$. It should be noted
that these correction terms dominate at large distances \cite{key-21}. 

Using the above mentioned modified Hawking temperature, the modified
form of the Boltzmann factor is 
\begin{equation}
\beta^{(m)}=\frac{1}{T_{H}^{(m)}}=\frac{1}{T_{H}\left(1-\frac{\alpha}{M^{2}}-\frac{\gamma}{M^{4}}\right)}=\frac{\beta_{H}}{\left(1-\frac{\alpha}{M^{2}}-\frac{\gamma}{M^{4}}\right)}.
\end{equation}
Thus, the (quantum corrected) modified BH mass has the expression
\begin{equation}
M^{(m)}=\frac{M}{\left(1-\frac{\alpha}{M^{2}}-\frac{\gamma}{M^{4}}\right)}.
\end{equation}
In case of emitted radiation from the BH, the modified Hawking temperature
(with quantum correction) becomes 
\begin{equation}
T_{H}^{(m)}=\frac{1}{8\pi M^{(m)}}.
\end{equation}
As a result, following \cite{key-23}, one can again use Hawking's
periodicity argument \cite{key-11,key-23,key-24} to obtain the modified
Schwarzschild like line element, which takes the form \cite{key-10,key-11}
\begin{equation}
\left(ds_{m}\right)^{2}=-(1-\frac{2M^{(m)}}{r})dt^{2}+\frac{dr^{2}}{1-\frac{2M^{(m)}}{r}}+r^{2}(\sin^{2}\theta d\varphi^{2}+d\theta^{2})
\end{equation}
with modified surface gravity 
\begin{equation}
\kappa{}^{(m)}=\frac{1}{4M^{(m)}}=\frac{1}{2r^{(m)}}=\frac{\left(1-\frac{\alpha}{M^{2}}-\frac{\gamma}{M^{4}}\right)}{4M}.
\end{equation}
Eq. (16) enables the replacement $M\rightarrow M^{(m)}$ and $T_{H}\rightarrow T_{H}^{(m)}$
in eqs. (3), (5)and (6). In other words, one can define the (quantum
corrected) \emph{modified effective temperature} 
\begin{equation}
T_{E}^{m}(\omega)\equiv\frac{2M^{(m)}}{2M^{(m)}-\omega}T_{H}^{(m)}=\frac{1}{4\pi(2M^{(m)}-\omega)},
\end{equation}
the (quantum corrected) \emph{modified effective Boltzmann factor
}
\begin{equation}
\beta_{E}^{(m)}(\omega)\equiv\frac{1}{T_{E}^{m}(\omega)}
\end{equation}
and the (quantum corrected) \emph{modified effective mass} and \emph{effective
horizon} radius 
\begin{equation}
M_{E}^{(m)}=M^{(m)}-\frac{\omega}{2}\qquad and\qquad r_{E}^{(m)}=2M_{E}^{(m)}=2M^{(m)}-\omega.
\end{equation}
A clarification is needed concerning the definition (18) \cite{key-28}.
Eqs. (14)-(17) give the quantum corrections using Hamilton-Jacobi
method beyond semiclassical approximation. Here we considered the
contributions of the non-thermal spectrum by reformulation of tunnelling
probability choosing Eq. (16) as the modified Schwarzschild line element. 
{\bf It should be noted that a full calculation involving the action of a 
particle on the BH spacetime also leads to this result.}
So Eq. (3) now becomes eq. (18). Following \cite{key-11,key-23,key-24},
one uses again Hawking's periodicity argument. Then, the euclidean
form of the metric will be given by 

\begin{equation}
\left[ds_{E}^{(m)}\right]^{2}=x^{2}\left[\frac{d\tau}{4M^{(m)}\left(1-\frac{\omega}{2M^{(m)}}\right)}\right]^{2}+\left(\frac{r}{r_{E}^{(m)}}\right)^{2}dx^{2}+r^{2}(\sin^{2}\theta d\varphi^{2}+d\theta^{2}),
\end{equation}
which is regular at $x=0$ and $r=r_{E}^{(m)}$. $\tau$ is treated
as an angular variable with period $\beta_{E}^{(m)}(\omega)$ \cite{key-11,key-23,key-24}.
Replacing the quantity $\sum_{i}\beta_{i}\frac{\hslash^{i}}{M^{2i}}$
in \cite{key-23} with the quantity $-\frac{\omega}{2M^{(m)}},$ if
one follows step by step the detailed analysis in \cite{key-23} at
the end one easily gets the (quantum corrected) \emph{modified effective
Schwarzschild line element 
\begin{equation}
\left[ds_{E}^{(m)}\right]^{2}=-(1-\frac{2M_{E}^{(m)}}{r})dt^{2}+\frac{dr^{2}}{1-\frac{2M_{E}^{(m)}}{r}}+r^{2}(\sin^{2}\theta d\varphi^{2}+d\theta^{2}).
\end{equation}
}One also easily shows that $r_{E}^{(m)}$ in eq. (21) is the same
as in eq. (20). Thus,the line element (22) takes into account both
the BH dynamical geometry during the emission of the particle and
the quantum corrections to the semiclassical Hawking temperature. 

Starting from the standard Schwarzschild line element, i.e. \cite{key-7,key-11}
\begin{equation}
ds^{2}=-(1-\frac{2M}{r})dt^{2}+\frac{dr^{2}}{1-\frac{2M}{r}}+r^{2}(\sin^{2}\theta d\varphi^{2}+d\theta^{2}),\label{eq: Hilbert}
\end{equation}
the analysis in \cite{key-7} permitted to write down the (normalized)
physical states of the system for bosons and fermions as \cite{key-7}
\begin{equation}
\begin{array}{c}
|\Psi>_{boson}=\left(1-\exp\left(-8\pi M\omega\right)\right)^{\frac{1}{2}}\sum_{n}\exp\left(-4\pi nM\omega\right)|n_{out}^{(L)}>\otimes|n_{out}^{(R)}>\\
\\
|\Psi>_{fermion}=\left(1+\exp\left(-8\pi M\omega\right)\right)^{-\frac{1}{2}}\sum_{n}\exp\left(-4\pi nM\omega\right)|n_{out}^{(L)}>\otimes|n_{out}^{(R)}>.
\end{array}\label{eq: physical states}
\end{equation}

Hereafter we focus the analysis only on bosons. In fact, for fermions
the analysis is identical \cite{key-7}. The density matrix operator
of the system is \cite{key-7}

\begin{equation}
\begin{array}{c}
\hat{\rho}_{boson}\equiv\Psi>_{boson}<\Psi|_{boson}\\
\\
=\left(1-\exp\left(-8\pi M\omega\right)\right)\sum_{n,m}\exp\left[-4\pi(n+m)M\omega\right]|n_{out}^{(L)}>\otimes|n_{out}^{(R)}><m_{out}^{(R)}|\otimes<m_{out}^{(L)}|.
\end{array}\label{eq: matrice densita}
\end{equation}
If one traces out the ingoing modes, the density matrix for the outgoing
(right) modes reads \cite{key-7}
\begin{equation}
\hat{\rho}_{boson}^{(R)}=\left(1-\exp\left(-8\pi M\omega\right)\right)\sum_{n}\exp\left(-8\pi nM\omega\right)|n_{out}^{(R)}><n_{out}^{(R)}|.\label{eq: matrice densita destra}
\end{equation}
This implies that the average number of particles detected at infinity
is \cite{key-7}

\begin{equation}
<n>_{boson}=tr\left[\hat{n}\hat{\rho}_{boson}^{(R)}\right]=\frac{1}{\exp\left(8\pi M\omega\right)-1},\label{eq: traccia}
\end{equation}
where the trace has been taken over all the eigenstates and the final
result has been obtained through a bit of algebra, see \cite{key-7}
for details. The result of eq. (\ref{eq: traccia}) is the well known
Bose-Einstein distribution. A similar analysis works also for fermions
\cite{key-7}, and one easily gets the well known Fermi-Dirac distribution

\begin{equation}
<n>_{fermion}=\frac{1}{\exp\left(8\pi M\omega\right)+1},\label{eq: traccia 2}
\end{equation}
Both the distributions correspond to a black body spectrum with the
Hawking temperature $T_{H}=\frac{1}{8\pi M}$. On the other hand,
if one follows step by step the analysis in \cite{key-7}, but starting
from the (quantum corrected) modified effective Schwarzschild line
element (22) at the end obtains the correct physical states for boson
and fermions as 
\begin{equation}
\begin{array}{c}
|\Psi>_{boson}=\left(1-\exp\left(-8\pi M_{E}^{(m)}\omega\right)\right)^{\frac{1}{2}}\sum_{n}\exp\left(-4\pi nM_{E}^{(m)}\omega\right)|n_{out}^{(L)}>\otimes|n_{out}^{(R)}>\\
\\
|\Psi>_{fermion}=\left(1+\exp\left(-8\pi M_{E}^{(m)}\omega\right)\right)^{-\frac{1}{2}}\sum_{n}\exp\left(-4\pi nM_{E}^{(m)}\omega\right)|n_{out}^{(L)}>\otimes|n_{out}^{(R)}>
\end{array}\label{eq: physical states-1}
\end{equation}
and the correct distributions as 
\begin{equation}
\begin{array}{c}
<n>_{boson}=\frac{1}{\exp\left(8\pi M_{E}^{(m)}\omega\right)-1}=\frac{1}{\exp\left[4\pi\left(2M^{(m)}-\omega\right)\omega\right]-1}=\frac{1}{\exp\left[4\pi\left(2\frac{M}{\left(1-\frac{\alpha}{M^{2}}-\frac{\gamma}{M^{4}}\right)}-\omega\right)\omega\right]-1}\\
\\
<n>_{fermion}=\frac{1}{\exp\left(8\pi M_{E}^{(m)}\omega\right)+1}=\frac{1}{\exp\left[4\pi\left(2M^{(m)}-\omega\right)\omega\right]+1}=\frac{1}{\exp\left[4\pi\left(2\frac{M}{\left(1-\frac{\alpha}{M^{2}}-\frac{\gamma}{M^{4}}\right)}-\omega\right)\omega\right]+1},
\end{array}\label{eq: final distributions}
\end{equation}
which are not thermal because they take into account both the BH dynamical
geometry during the emission of the particle and the quantum corrections
to the semiclassical Hawking temperature. We note that setting $\alpha=\gamma=0$
in eqs. (30) we find the results in \cite{key-11}, i.e. eqs. (8).
In fact, in \cite{key-11} only the BH dynamical geometry was taken
into account. Here, we further improved the analysis by taking into
account also the quantum corrections to the semiclassical Hawking
temperature.

\section*{Concluding remarks}

The present work deals with the quantum correction of non-thermal
radiation spectrum in the framework of tunnelling mechanism. Staring
from the Schwarzschild BH, at first the quantum corrections are considered.
As a result, the Hawking temperature and Schwarzschild mass are modified
(see Eqs. (14) and (15)). So one obtains the modified Schwarzschild
line element (see Eq. (16)). Then we consider the non-thermal radiation
spectrum of this modified Schwarzschild BH by the reformulation of
the tunnelling probability. The resulting quantum corrected effective
Schwarzschild metric is rewritten using Hawking's periodicity arguments.
{\bf Also we have shown the correct distributions of bosons and fermions
using the above quantum corrections to the semiclassical Hawking temperature.
Thus due to quantum correction at the semiclassical level, BH parameters 
(and its radiation spectrum), namely, its mass, temperature, surface gravity,
and Boltzmann factors are modified and as a result, we have quantum corrected
effective Schwarzschild metric. Moreover, the one-loop correction which comes 
from interaction between graviton and particles of various species (characterized)
in Eq. (11) occurred at the horizon. Hence, the quantum effects lead to a 
redefinition of surface gravity and other parameters. But it should be 
noted that the BH's gravitational potential may not only be characterized
by this modified mass far away from the horizon. Therefore, the modified 
metric [in Eq. (16) or Eq. (22)] can only be trusted for its near-horizon geometry,
but nowhere else--and the effective metric for arbitrary distance could be
elaborated in a perturbative way \cite{key-29, key-30} and the potential is not Coloumb-like in general.}

\section*{Acknowledgments}

The authors S. C. and S. S. are thankful to IUCAA, Pune, India for
their warm hospitality and research facilities as the work has been
done there during a visit. Also S. C. acknowledges the UGC-DRS Programme
in the Department of Mathematics, Jadavpur University. The author
S. S. is thankful to UGC-BSR Programme of Jadavpur University for
awarding research fellowship. The authors thank the unknown referees
for important advices which permitted to improve this work.

\end{document}